\documentclass[11pt,twoside]{article}
\usepackage{asp2010}

\resetcounters
\aspvolume{455}
\aspcpryear{2012}
\aspvoltitle{4$^{th}$ {\em Hinode} Science Meeting: Unsolved 
Problems and Recent Insights}
\aspvolauthor{L.~R.~Bellot Rubio, F.~Reale, and M.~Carlsson, eds.}

\begin{document}

\title{A Multi-Wavelength Statistical Study of Supra-Arcade Downflows}
\author{Sabrina L.~Savage and David E.~McKenzie}
\affil{Department of Physics, Montana State University, 
P.O.~Box 173840, Bozeman, MT 59717-3840, USA}

\begin{abstract}
Sunward-flowing voids above post-coronal mass ejection flare arcades,
also known as supra-arcade downflows (SADs), have characteristics
consistent with post-reconnection magnetic flux tube
cross-sections. Applying semi-automatic detection and analysis
software to a large sample of flares using several instruments
(e.g., \textit{Hinode}/XRT, \textit{Yohkoh}/SXT, TRACE, and
\textit{SOHO}/LASCO), we have estimated parameters such as speeds,
sizes, heights, magnetic flux, and relaxation energy associated with
SADs, which we interpret as reconnection outflows. We also present
speed and height measurements of shrinking loops in
comparison to the SAD observations.  We briefly discuss these
measurements and what impact they have on reconnection models.
\end{abstract}

\vspace*{-.7em}

\section{Introduction}

\begin{figure}[!t] 
\centering 
\includegraphics[width=.55\textwidth]{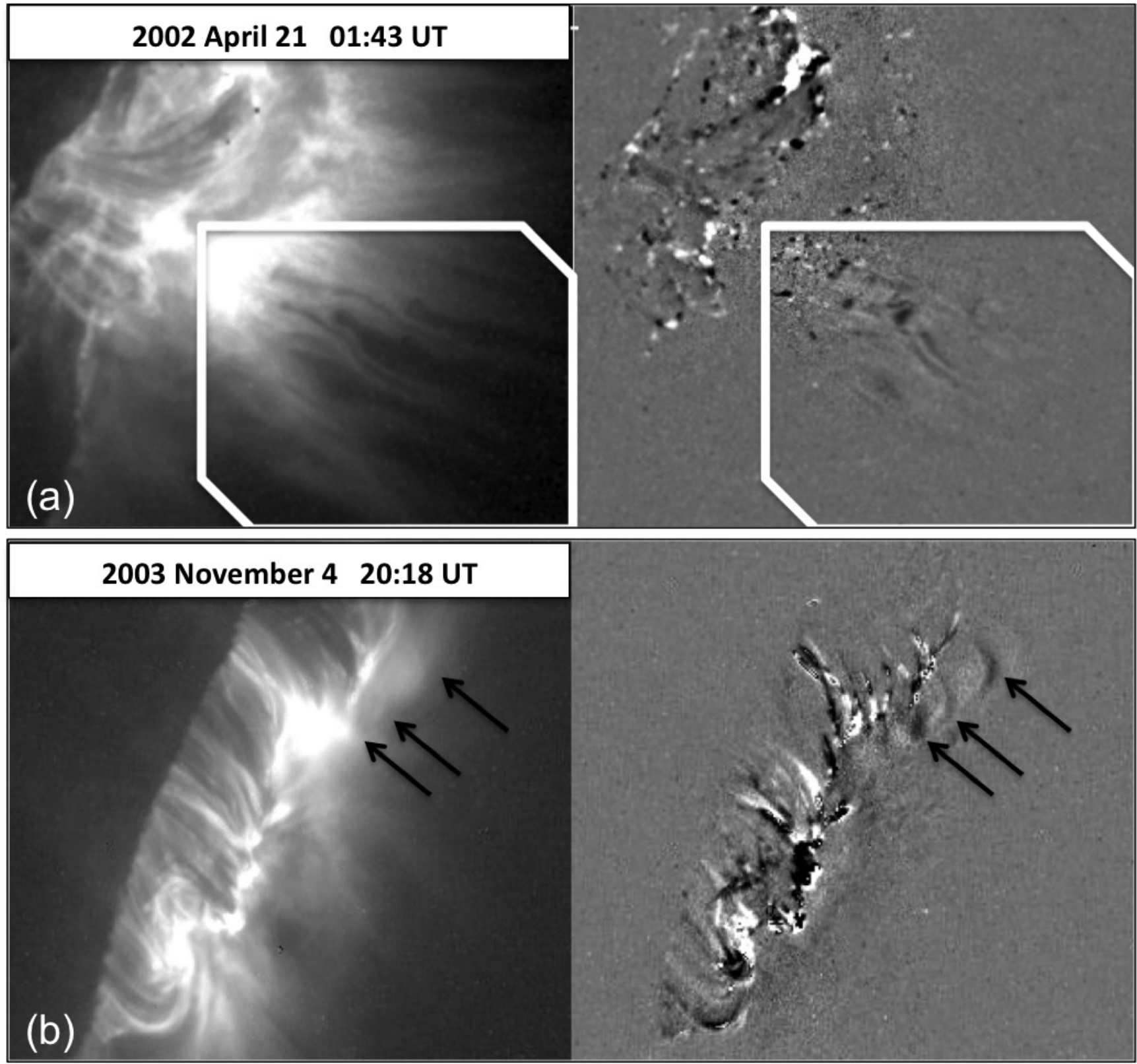}
\caption{{\em (a)} Example image from the April 21, 2002 TRACE 
flare showing supra-arcade downflows (SADs) enclosed within the white
box.  {\em (b)} Example image from the November 4, 2003 flare with
supra-arcade downflowing loops (SADLs) indicated by the arrows.  The
left panel of each set is the original image.  The right panel has
been enhanced for motion via run-differencing and scaled for
contrast.}
\label{sads_sadls_example}
\end{figure}

Long duration flaring events are often associated with downflowing
voids and/or loops in the supra-arcade region (see
Fig.~\ref{sads_sadls_example} for example images) whose theoretical
origin as newly reconnected flux tubes has been supported by
observations \citep{1999ApJ...519L..93M,2000SoPh..195..381M,
2003SoPh..217..247I,2004ApJ...605L..77A,2004ApJ...616.1224S, 
2007A&A...475..333K, 2008ApJ...675..868R, 2009ApJ...697.1569M,
2010ApJ...722..329S}.

\begin{figure}[!ht]
\vspace*{1em} 
\begin{center}
\includegraphics[width=0.9\textwidth]{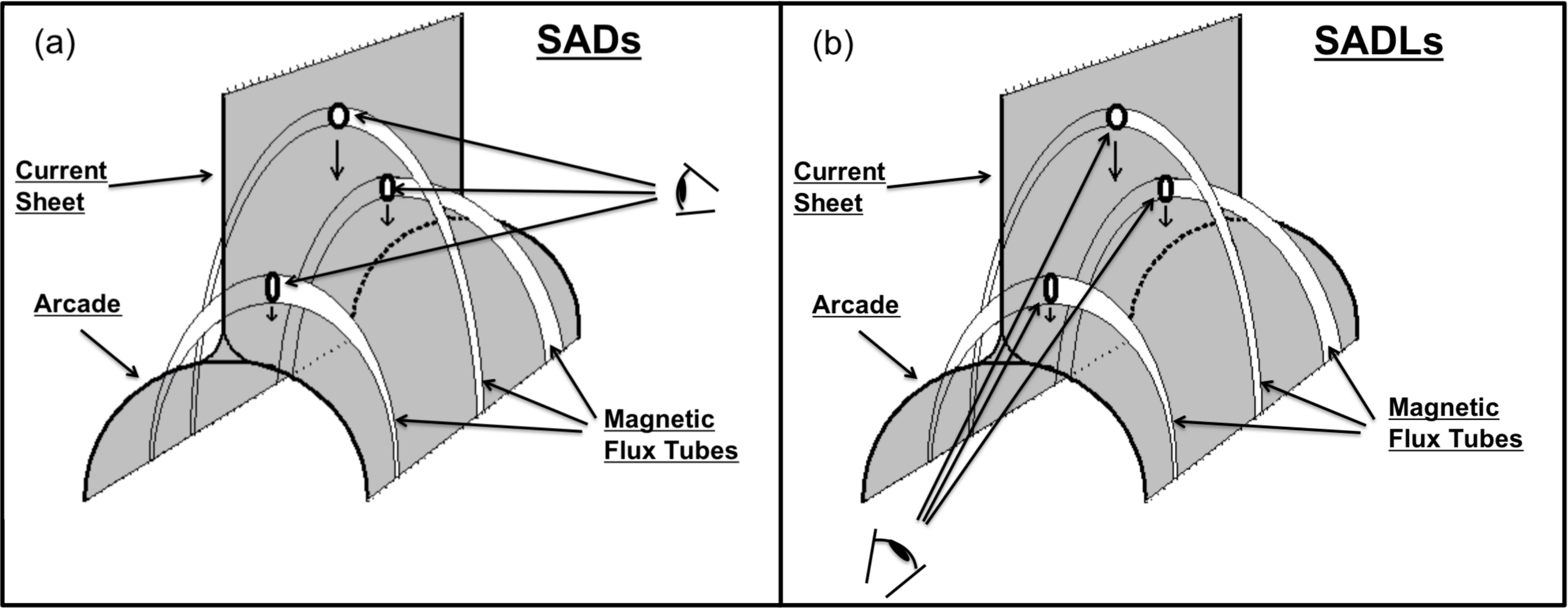}
\caption{{\em (a)} Cartoon depiction of supra-arcade downflows (SADs) 
resulting from 3D patchy reconnection. Discrete flux tubes are
created, which then individually shrink, dipolarizing to form the
post-eruption arcade. {\em (b)} Cartoon depiction of supra-arcade
downflowing loops (SADLs) also resulting from 3D patchy reconnection.
Note that the viewing angle, indicated by the eye position, is
perpendicular to that of SADs observations.}
\label{sads_sadls_diagram_eyes_ch4}
\end{center}
\end{figure}

The downflowing voids, or supra-arcade downflows (SADs;
Fig~\ref{sads_sadls_example}a), differ in appearance from downflowing
loops, or supra-arcade downflowing loops (SADLs;
Fig.\ \ref{sads_sadls_example}b); however, the explanation for this can
be derived simply from observational perspective.  If the loops are
viewed nearly edge-on as they retract through a bright current sheet,
then SADs may represent the cross-sections of the SADLs (see
Figure~\ref{sads_sadls_diagram_eyes_ch4}).  Since neither SADs nor
SADLs can be observed 3-dimensionally by an independent imaging
instrument, proving this hypothetical connection is not possible with
a single image sequence.  However, their general bulk properties, such
as velocity, size, and magnetic flux, can be measured and should be
comparable if this scenario is correct.  Moreover, measuring these
parameters for a large sample of SADs and SADLs yields constraints
that are useful for development of numerical models/simulations of 3D
magnetic reconnection in the coronae of active stars.  We present
analysis of flows from 35 flares and compare the results of general
bulk properties, including magnetic flux and shrinkage energy
estimates, from SADs and SADLs.  These comparisons provide compelling
evidence linking SADs to SADLs and constraints on flare magnetic
reconnection models.

\section{\label{sadsiisec:analysis}Analysis}

Considering the substantial uncertainty sources associated with flow
detections, flow measurements should be taken as imprecise; however,
the large number of fairly well-defined limb downflows (total of 369)
tracked from our flare list make it possible to consider ranges and
trends in the data.

\subsection{\label{sadsiisec:quartiles}Synthesis of Frequency Diagrams}

\begin{figure}[!p] 
\begin{center}
\includegraphics[width=0.96\textwidth,bb=20 0 734 361]{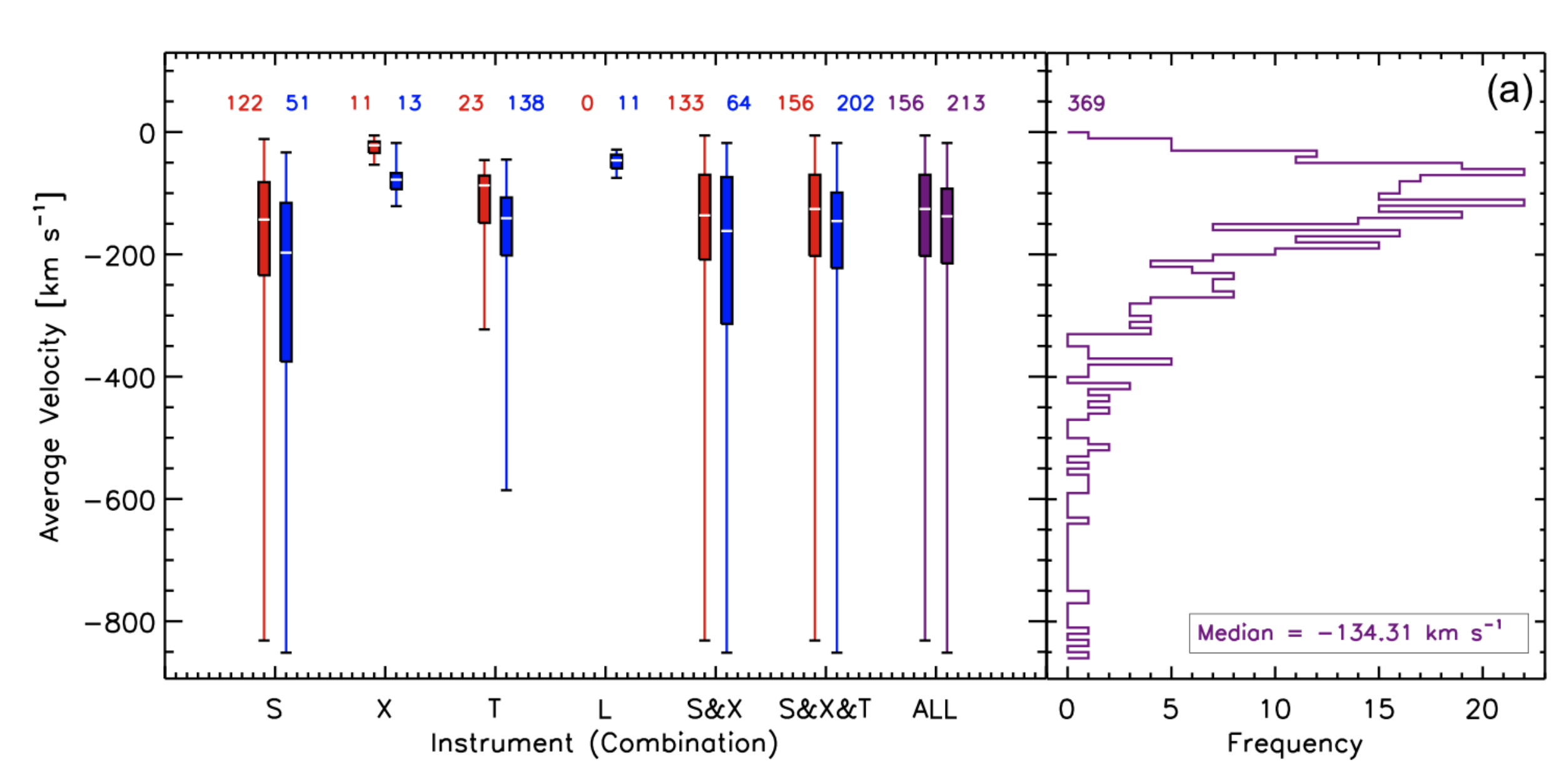}
\includegraphics[width=0.96\textwidth,bb=20 0 734 361]{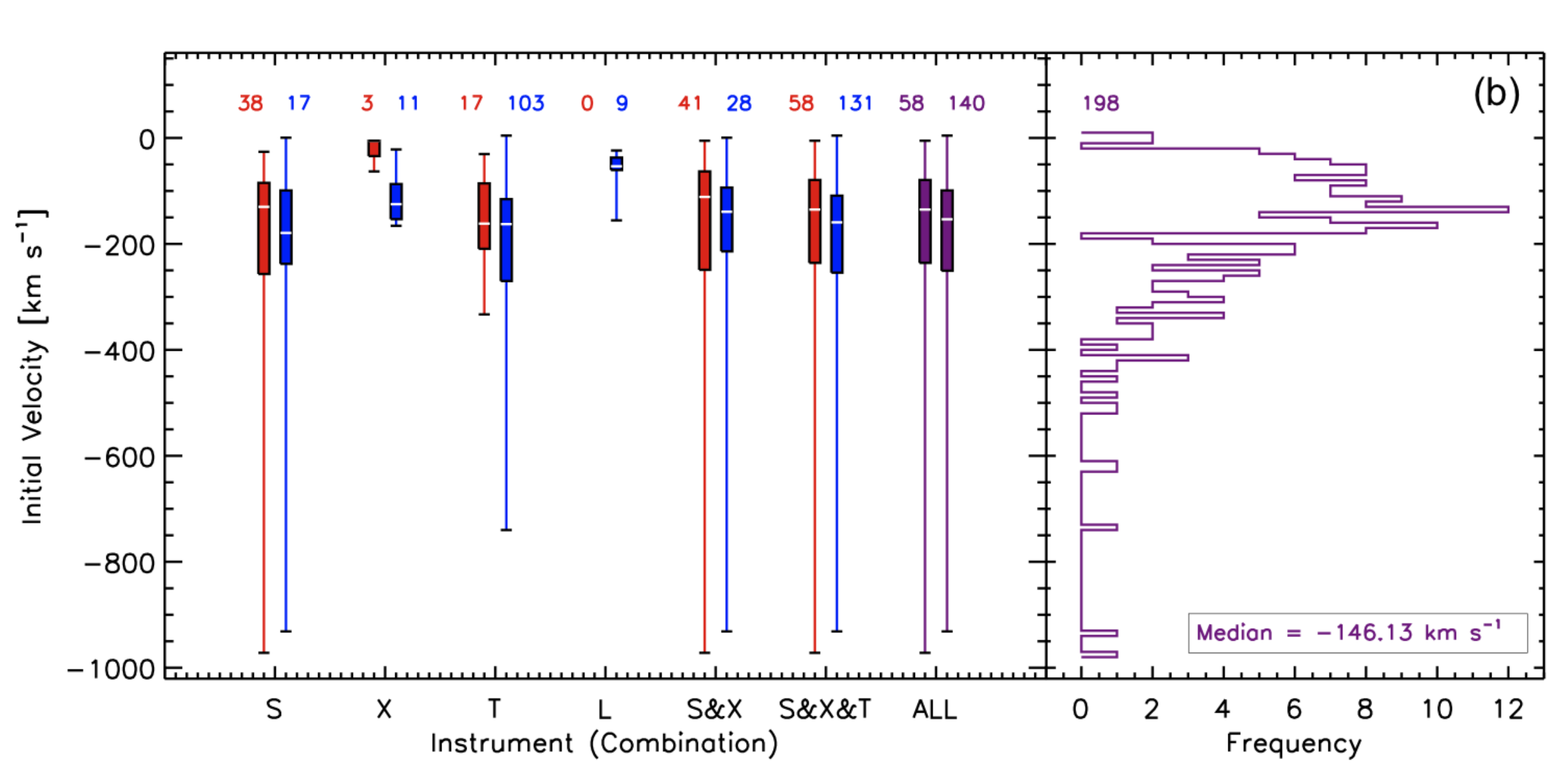}
\includegraphics[width=0.96\textwidth,bb=20 0 734 361]{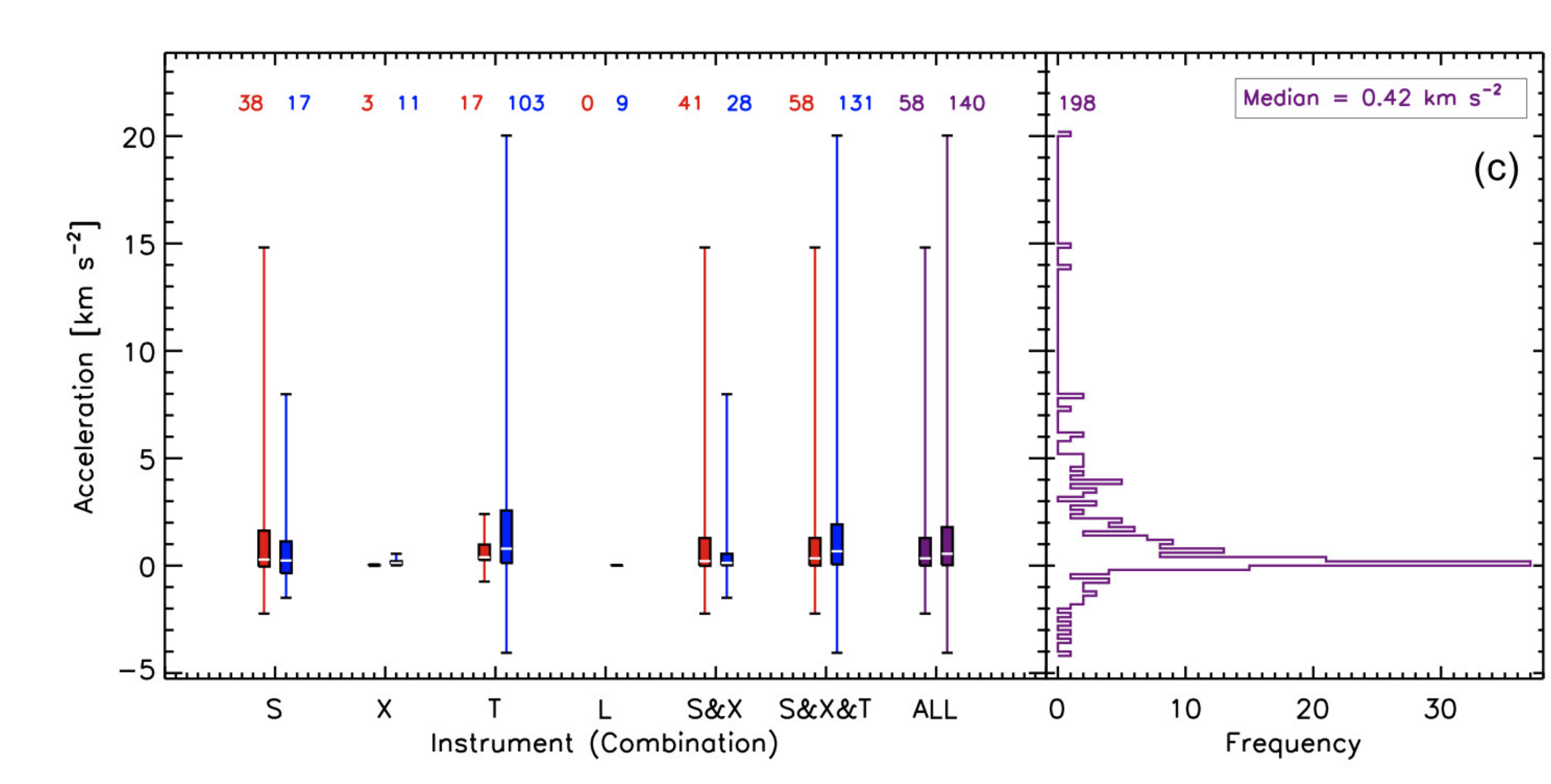}
\caption{Synthesis of the de-projected trajectory parameter estimates. 
{\em (a)} De-projected average velocity.  {\em (b)} De-projected
initial velocity. {\em (c)} De-projected acceleration. (Refer to the
text for a detailed description of these figures.)}
\label{quart1}
\end{center}
\end{figure}

Figures~\ref{quart1},~\ref{quart2}, and~\ref{quart3} synthesize the
flow measurement results from all of the flares under consideration.
Each plot in the figures consists of a quartile plot in the left
panel and a histogram in the right.  For the quartile plots, the
measurement is plotted against the instrument (or instrument
combination) being considered (S: SXT; X: XRT; T: TRACE; L: LASCO;
All: S\&X\&T\&L).  For Figs.~\ref{quart1} and~\ref{quart2}, the left
(red or purple) box-and-whisker range per instrument (or instrument
combination) represents SADs measurements while the right (blue or
purple) one represents SADLs measurements.  For Fig.~\ref{quart3}, the
east (pink or green) and west (olive or green) limbs are compared
instead of SADs to SADLs.  The lines (or whiskers) extending from the
boxes indicate the full range of the data.  The boxes span the range
of the middle 50\% of the data.  The (white) line through the box
indicates the median of the data.  Along the top of these plots, the
number of flows used to derive the associated measurements is labeled.
The combination of the data in the final two (purple or green)
box-and-whisker plots is contained within the histogram panel.  The
median of the histogram is displayed in the legend.

LASCO measurements are not included in Figs.~\ref{quart2}
or~\ref{quart3} since its resolution (11.4~arcsec/pix) is so much
poorer and its observational regime high above the limb
($>$~2.5~R$_{\odot}$ for C2) is so very different from that of the
other instruments, making comparisons more complicated.  Deriving
magnetic fields at such heights is not applicable with our method
either plus determining precise footpoints without coincidental data
from other instruments is nearly impossible.  The total number of
flows under consideration after removing those observed by LASCO is
358.

\subsubsection{Velocity and Acceleration}

There is general agreement between SADs and SADLs, the instruments,
and the SXR versus EUV bandpasses for the average velocity, initial
velocity, and acceleration measurements (Figure~\ref{quart1}).  Note
that the initial velocity and acceleration plots do not incorporate
all 369 available flows.  Instead, only those flows tracked in at
least 5 frames were included because these measurements rely on
fitting the trajectories to a 2D polynomial fit.  Using fewer than 5
points leads to unreliable results.  Also note that a positive
downflow acceleration means that the flow is slowing.

\subsubsection{Area}

\begin{figure}[!p] 
\begin{center}
\includegraphics[width=0.96\textwidth,bb=20 0 734 361,clip]{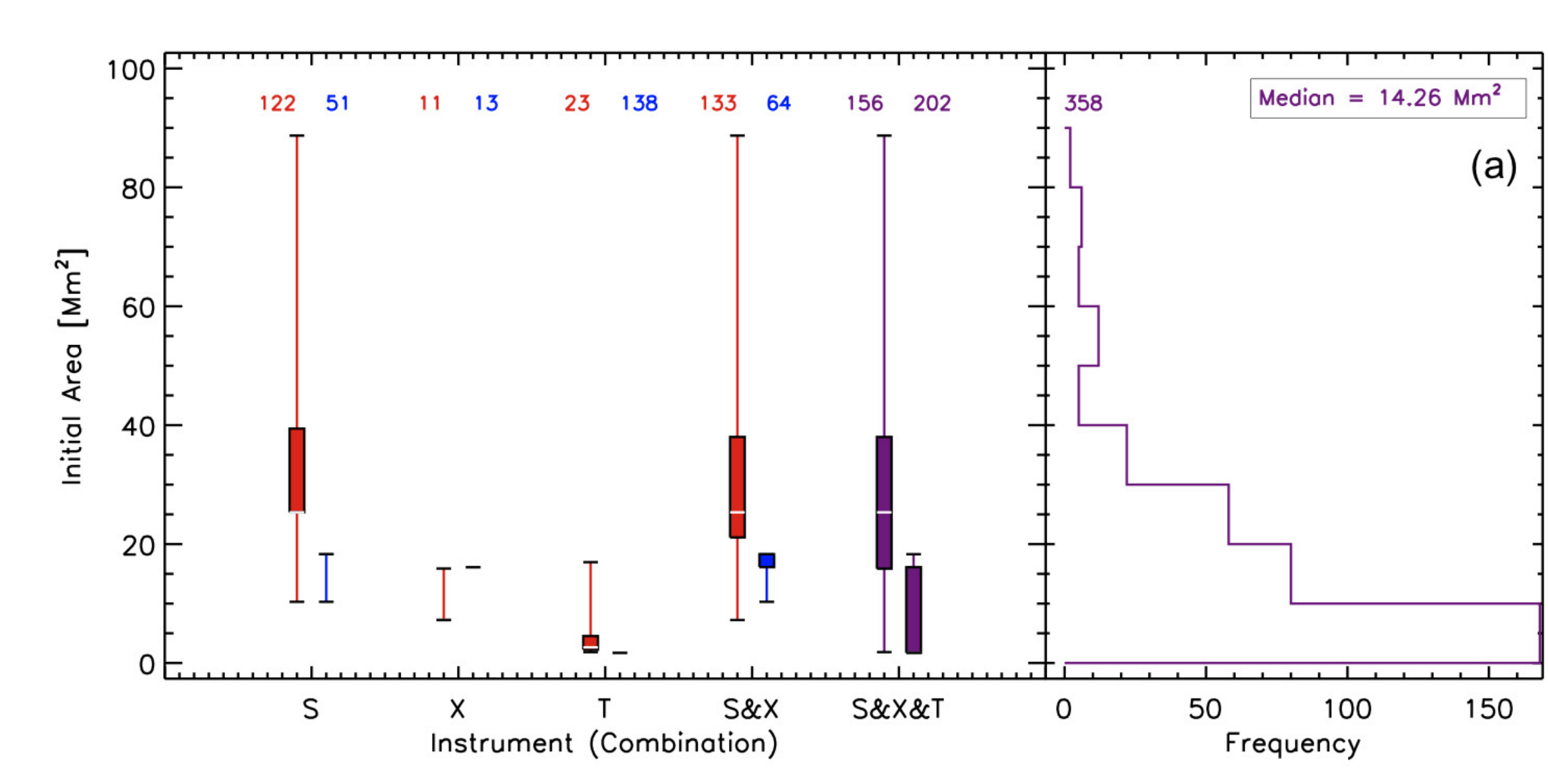}
\includegraphics[width=0.96\textwidth,bb=20 0 734 361,clip]{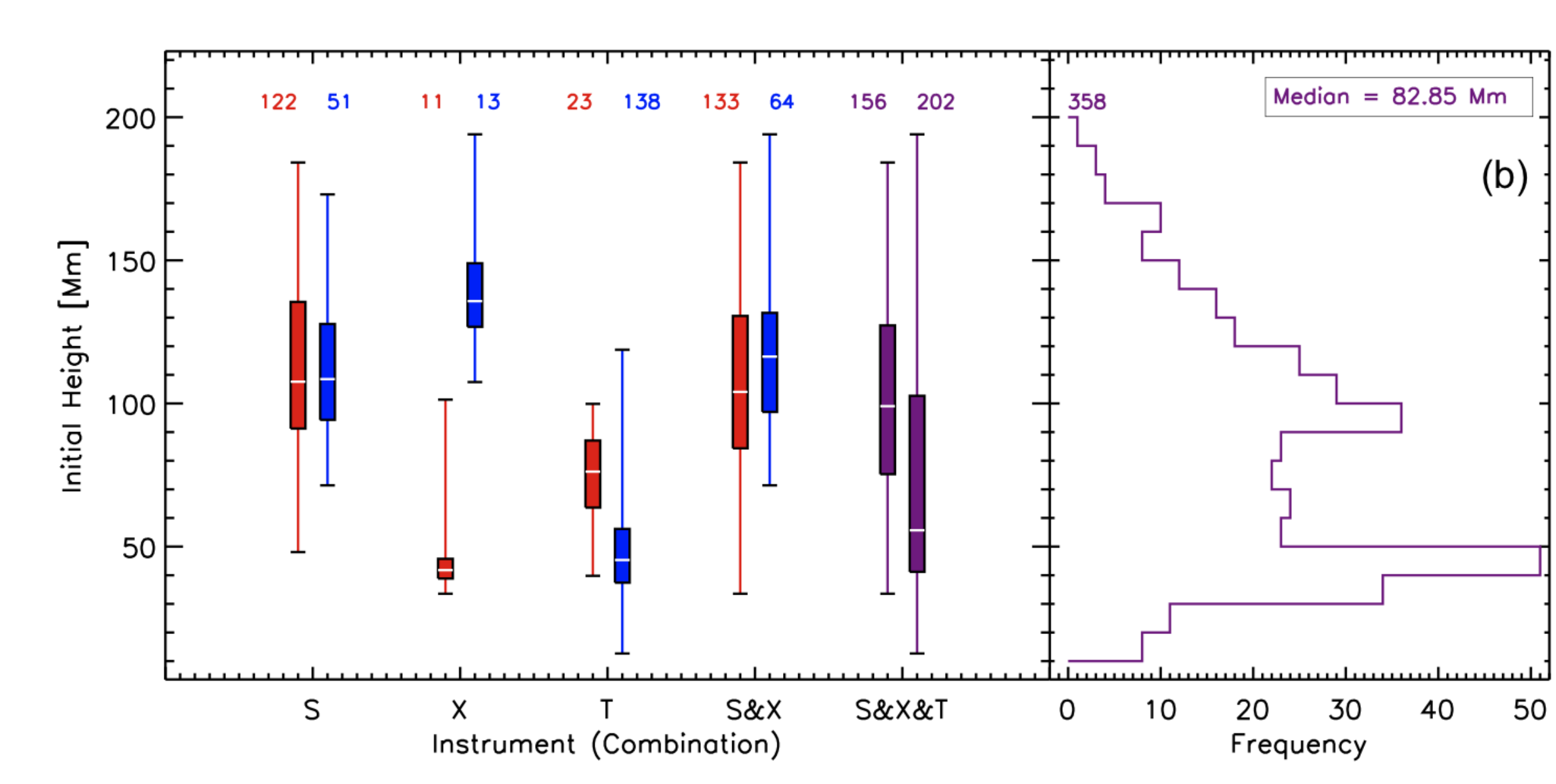}
\includegraphics[width=0.96\textwidth,bb=20 0 734 361,clip]{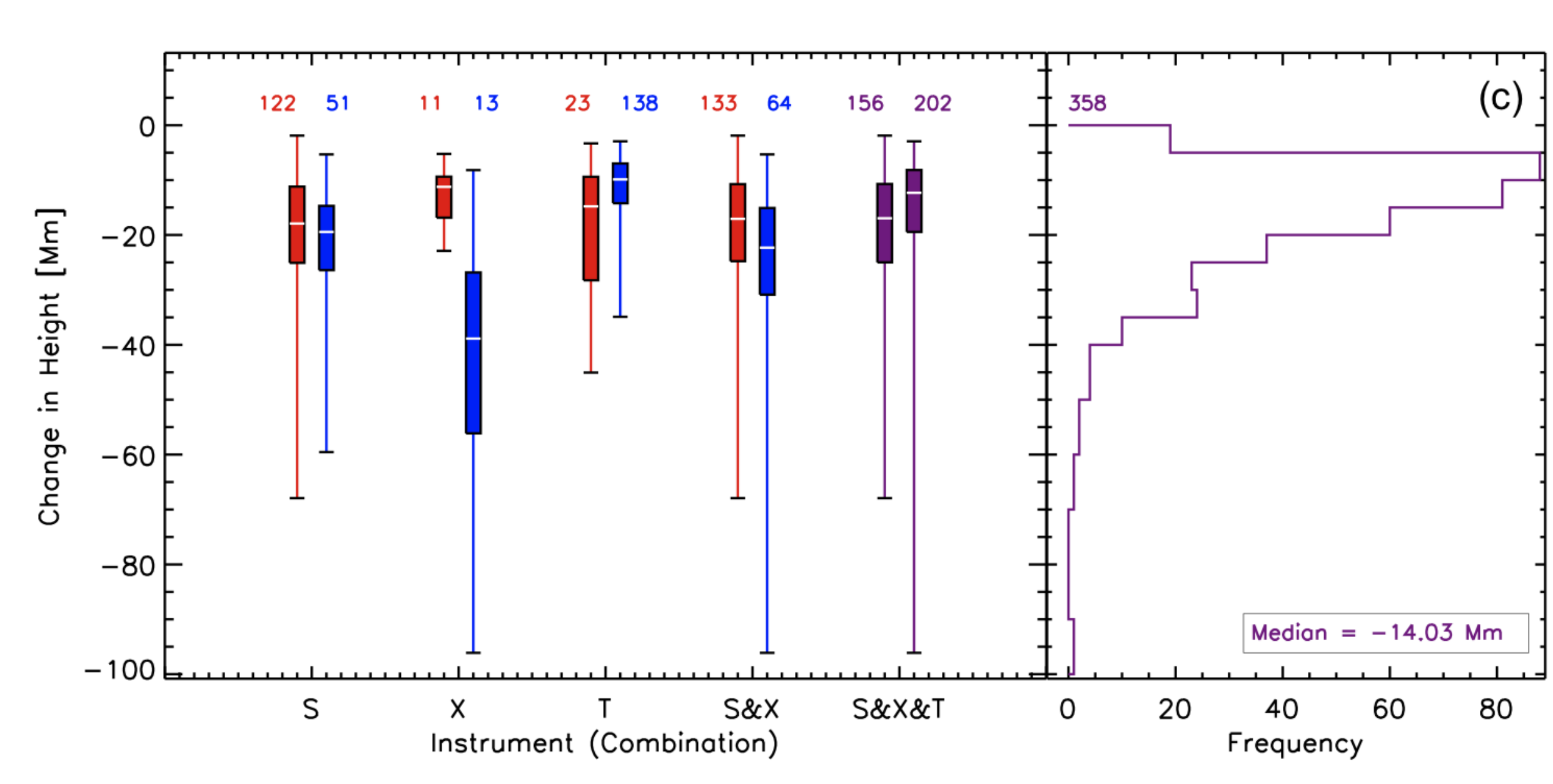}
\caption{Synthesis of the area and height parameter estimates.  
{\em (a)} Initial area.  {\em (b)} De-projected initial height.  {\em
(c)} De-projected change in height. (Refer to the text for a detailed
description of these figures).}
\label{quart2}
\end{center}
\end{figure}

A strong correspondence between instrument resolution (SXT:
2.5--4.9~arcsec/pix; XRT: 1~arcsec/pix; TRACE: 0.5 arcsec/pix) and
measured area is shown in the initial area quartile plot
(Figure~\ref{quart2}a).  The SADLs and XRT SADs area measurements are
very strongly peaked due to their manual assignments (versus threshold
detection)---hence the lack of distinct quartiles.

\begin{figure}[!p] 
\begin{center}

\includegraphics[width=0.96\textwidth,bb=20 0 734 361]{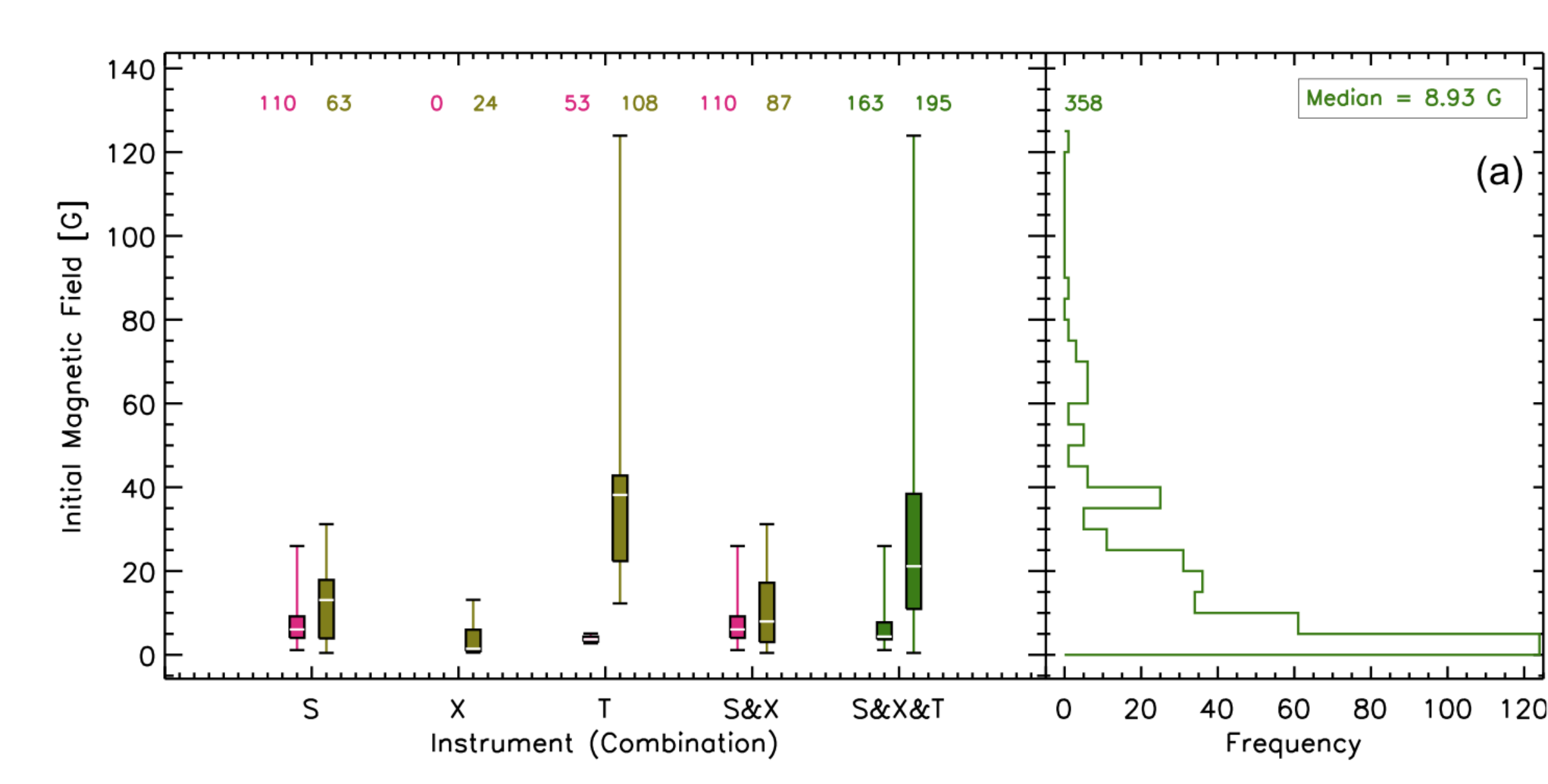}
\includegraphics[width=0.96\textwidth,bb=20 0 734 361]{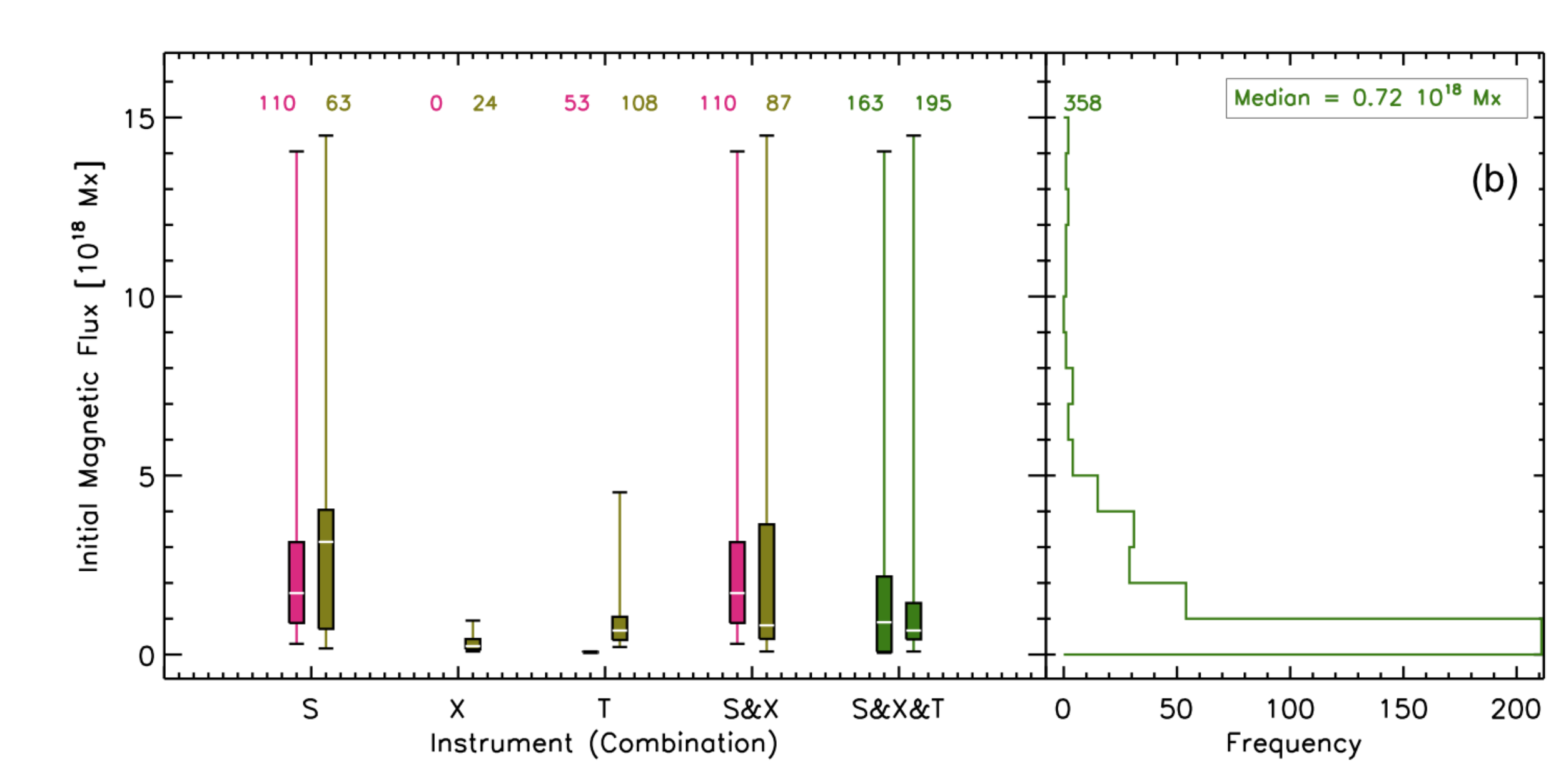}
\includegraphics[width=0.96\textwidth,bb=20 0 734 361]{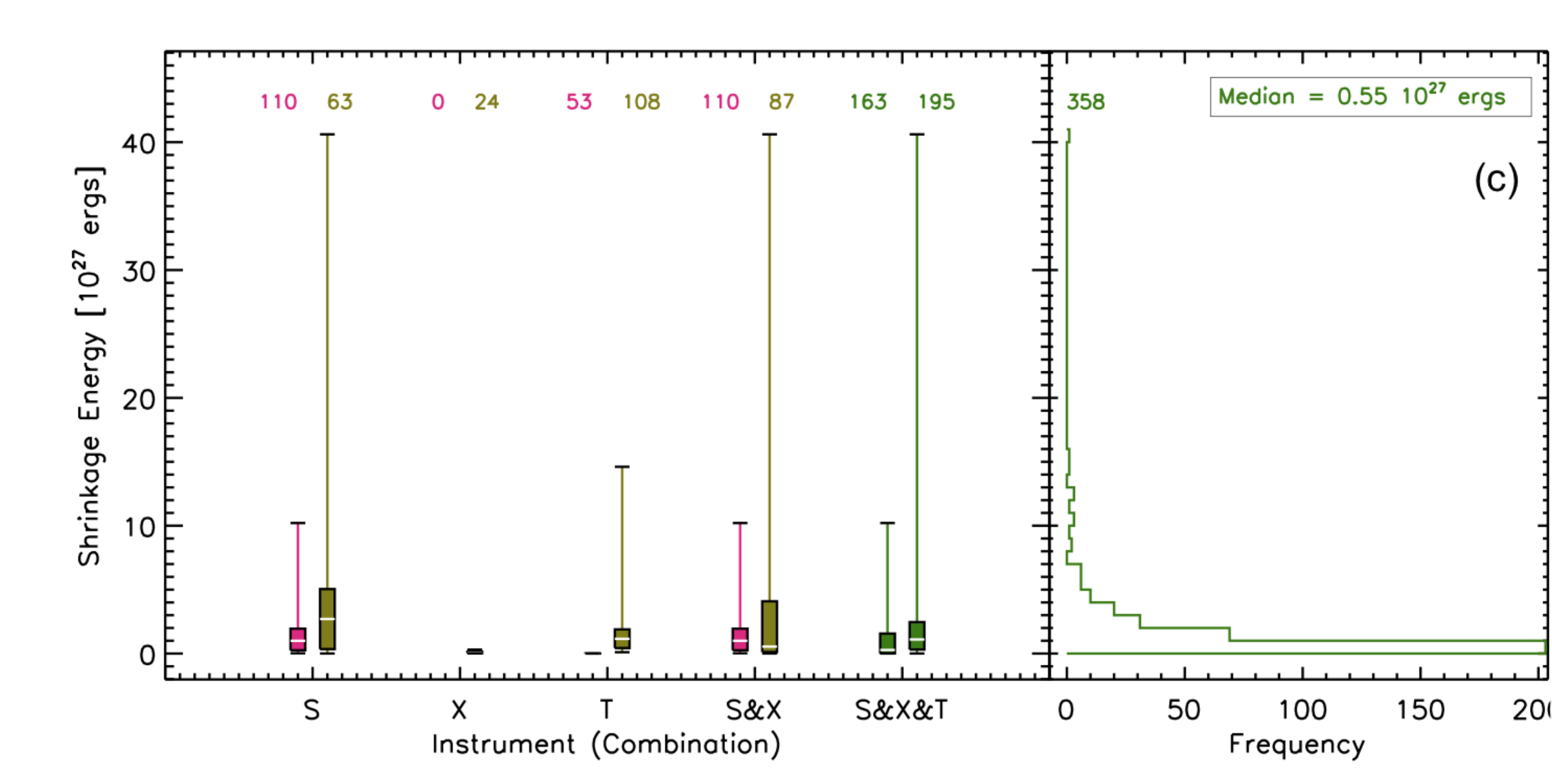}

\caption{Synthesis of the magnetic parameter estimates  {\em (a)}  
Initial magnetic field.  {\em (b)} Initial magnetic flux.  {\em (c)}
Shrinkage energy. (Refer to the text for a detailed description of
these figures).}
\label{quart3}
\end{center}
\end{figure}

\subsubsection{Height}

The initial height ranges (Fig.~\ref{quart2}b) show decent agreement
between the ranges; however, there is a fair amount of scatter in the
medians which requires more detailed understanding of the analyzed
flares to explain.  The initial heights for both SADs and SADLs
observed by SXT offer very good agreement.  XRT observations, while
agreeing with SXT's range of initial heights, show no agreement
between SADs and SADLs.  This discrepancy is due to a combination of
factors: (1) XRT has observed very few SADs as yet; (2) XRT
observations are rarely sufficiently exposed to illuminate the
supra-arcade region; therefore, XRT SADs have only been observed
nearer to the solar surface; (3) The SADLs observed by XRT are derived
from the ``Cartwheel CME" flare \citep{2010ApJ...722..329S} during
which the footpoints were obscured by the limb enabling very long
exposure durations.  In fact, a disconnection event associated with
this flare (\cite{2010ApJ...722..329S}, Sect.~3.2 therein) was
observed at nearly 190~Mm above the solar surface, which is at the
maximum of the combined instrument ranges.  TRACE's temperature
coverage targets plasma on order of 1~MK with some overlap in the
11--26~MK range with the SXR imagers.  The image exposure durations
are also optimized to observe the flaring region near the solar
surface. Consequently, the observed initial heights of SADs and SADLs
measured with TRACE are limited to the region near the top of the
growing post-eruption arcade where the hot plasma in the current sheet
is most illuminated.  This results in initial heights lower than many
of those reported for SXT and XRT.

The change in heights shown in Fig.~\ref{quart2}c are naturally flare
and field of view (FOV) dependent.  Even so, there is general
agreement between SADs, SADLs, and instrument except for XRT.  The
explanation for this XRT discrepancy is the same as that for the
initial height XRT discrepancy described above (i.e., the flows for
the ``Cartwheel CME" flare could be tracked further through the FOV).

\subsubsection{Magnetic Measurements}

Figure~\ref{quart3}a is provided as a visual reference for the initial
magnetic fields which are used to calculate the magnetic flux ($\Phi =
B \times A$, Fig.~\ref{quart3}b) and the shrinkage energy ($\Delta W =
B^2 A \Delta L / 8\pi$, Fig.~\ref{quart3}c). Refer to 
\cite{2009ApJ...697.1569M}, Sect.~4.5, for a detailed description 
of the shrinkage energy calculations. The initial magnetic field
estimates for the TRACE flares are larger than the majority of those
from SXT and XRT because (1) the TRACE flares analyzed are highly
energetic according to their GOES classifications and (2) the flows
are observed closer to the surface (see Fig.~\ref{quart2}b) where the
magnetic field is stronger according to the PFSS model
\citep{1969SoPh....6..442S,2003SoPh..212..165S}.
East and west limb measurements are compared in Fig.~\ref{quart3} to
show the effect of using less reliable east-limb magnetograms.  The
tendency for west limb flares to have stronger initial magnetic field
estimates is noticeable in Fig.~\ref{quart3}a and carries through into
the initial magnetic flux and shrinkage energy plots
(Figs.~\ref{quart3}b and \ref{quart3}c, respectively).  The dichotomy
is most noticeable in the shrinkage energy estimates due to the $B^2$
component.

\subsection{Discussion}

Interpreting SADs as the cross-sections of retracting reconnected flux
tubes also means that if they are viewed from an angle that is not
near perpendicular to the arcade axis (i.e., the polarity inversion
line), the downflows will instead appear as shrinking loops.  These
shrinking loops (SADLs) have indeed been clearly observed with all of
the instruments under investigation.  Therefore, comparing
observations of SADs to those of SADLs can help to support or refute
the SADs hypothesis.

Figures~\ref{quart1} through \ref{quart3} present a summary of the
instrument and SAD/SADL comparisons.  These figures show that the flow
velocities and accelerations agree between the instruments quite well.
Height measurements agree except for those measured with XRT due to
the exceptional flow heights observed for the ``Cartwheel CME" flare
\citep{2010ApJ...722..329S}.  Figure~\ref{quart2}a shows that the
area measurements are understandably resolution dependent, which
indicates that we may not be able to observe the smallest loop sizes.
The flux and energy measurements are area dependent and therefore
instrument dependent.  There is also a limb dependence with the
magnetic measurements due to the use of modeling based on
magnetograms.  Even so, there is decent agreement between all of the
instruments (LASCO is only included with the velocity and acceleration
comparisons as explained in Section~\ref{sadsiisec:quartiles}).
Beyond the agreement between the SADs and SADLs measurements, the
high-resolution TRACE observations clearly show both SADs and SADLs
occurring during the same flare depending on the arcade viewing angle
which curves within the active region.

The measured cross-sectional areas range from $\sim$2 to 90~Mm$^{2}$,
with at least 75\% being smaller than 40~Mm$^{2}$
(Figure~\ref{quart2}a).  The flows typically move at speeds on
order of 10$^{2}$~km~s$^{-1}$ with accelerations that are near zero or
slightly decelerating.  The most complete flow paths show significant
deceleration near the top of the arcade.  There is a range of initial
heights depending on the quality of the image set, but they are
generally about 10$^{5}$~km above the solar surface with a path length
of $\sim$10$^{4}$~km.  Each tube carries $\sim$10$^{18}$~Mx of flux
and releases on order of 10$^{27}$~ergs of energy as it retracts.  A
lower limit of 10$^{16}$~Mx~s$^{-1}$ can be put on the reconnection
rate by considering the total flux released by the observed flows for
5 flares from our list.

The observational findings presented here provide a more complete
description of the SAD/SADL phenomenon than has previously been
available.  Assuming that SADs and SADLs are thin, post-reconnection
loops based on this body of evidence, the measurements obtained
through this analysis and summarized above provide useful constraints
for reconnection models.  Area estimates can constrain the diffusion
time per episode and reconnection rates can be derived to distinguish
between fast and slow reconnection.  Creation of outflowing flux tubes
carrying on order of 10$^{18}$~Mx of flux, with net reconnection rates
of at least 10$^{16}$~Mx~s$^{-1}$, should be an objective of realistic
models of 3D reconnection.  The lack of acceleration of the downflow
speeds and their discrete nature tends to favor 3D patchy Petschek
reconnection.  Speeds almost an order of magnitude slower than
traditionally assumed Alfv\'{e}n speeds are an unexpected consequence
of the flow measurements; therefore, analyzing the effect of some
source of drag on the downflow trajectories using models (an effort
begun by \citealt{2006ApJ...642.1177L}) could provide valuable insight
into this discrepancy.

\acknowledgements 

This work was supported by NASA grant NNM07AB07C.  SLS is grateful for
the meeting registration support provided by the {\em Hinode}-4 SOC.

\bibliography{hinode4}

\end{document}